\documentclass[twocolumn]{aastex631}
\usepackage{graphicx} 
\usepackage{threeparttable}

\begin{document}

\title{GRB 221009A: revealing a hidden afterglow during the prompt emission phase with  Fermi-GBM observations}

\author[0000-0001-6863-5369]{Hai-Ming Zhang}
\affil{School of Astronomy and Space Science, Nanjing University, Nanjing 210023, China, hmzhang@nju.edu.cn, ryliu@nju.edu.cn, xywang@nju.edu.cn}
\affil{Key laboratory of Modern Astronomy and Astrophysics (Nanjing University), Ministry of Education, Nanjing 210023, China}
\author[0000-0002-6036-985X]{Yi-Yun Huang}
\affil{School of Astronomy and Space Science, Nanjing University, Nanjing 210023, China, hmzhang@nju.edu.cn, ryliu@nju.edu.cn, xywang@nju.edu.cn}
\affil{Key laboratory of Modern Astronomy and Astrophysics (Nanjing University), Ministry of Education, Nanjing 210023, China}
\author[0000-0003-1576-0961]{Ruo-Yu Liu}
\affil{School of Astronomy and Space Science, Nanjing University, Nanjing 210023, China, hmzhang@nju.edu.cn, ryliu@nju.edu.cn, xywang@nju.edu.cn}
\affil{Key laboratory of Modern Astronomy and Astrophysics (Nanjing University), Ministry of Education, Nanjing 210023, China}
\author[0000-0002-5881-335X]{Xiang-Yu Wang}
\affil{School of Astronomy and Space Science, Nanjing University, Nanjing 210023, China, hmzhang@nju.edu.cn, ryliu@nju.edu.cn, xywang@nju.edu.cn}
\affil{Key laboratory of Modern Astronomy and Astrophysics (Nanjing University), Ministry of Education, Nanjing 210023, China}

\received{ 2023 July 24}
\revised{ 2023 September 17}
\accepted{2023 September 22}
\submitjournal{ApJ Letters}

\shorttitle{GRB 221009A}
\shortauthors{ZHANG et al.}

\begin{abstract}

Recently, LHAASO reported the detection of brightest-of-all-time GRB 221009A, revealing the early onset of a TeV afterglow. We analyze the spectral evolution of the  X-ray/$\gamma$-ray emission of GRB 221009A measured by Fermi Gamma-Ray Burst Monitor (GBM) during the dips of two prompt emission pulses (i.e., intervals $T_{0}+[300-328]\rm~s$ and $T_{0}+[338-378]\rm~s$, where $T_0$ is the GBM trigger time). We find that the spectra at the dips transit from the Band function to a power-law function, indicating a transition from the prompt emission to the afterglow. After $\sim T_{0}+ 660 \rm~s$, the spectrum is well described by a power-law function and the afterglow becomes dominant. Remarkably, the underlying afterglow emission at the dips smoothly connect with the  afterglow after $\sim T_{0}+ 660 \rm~s$. The entire afterglow emission measured by GBM can be fitted by a power-law function  $F\sim t^{-0.95\pm0.05}$, where $t$ is the time since the first main pulse at $T^*=T_0+226~{\rm s}$, consistent with the TeV afterglow decay measured by LHAASO. The start time of this power-law decay  indicates that the afterglow peak  of GRB 221009A should be earlier than $T_{0}+300 \rm ~s$. We also test the possible presence of a jet break in the early afterglow light curve, finding that both the jet break model and single power-law decay model are consistent with the GBM data. The two models can not be distinguished with the GBM data alone because the  inferred jet break time is quite close to the end of GBM observations.

\end{abstract}

\keywords{Gamma-ray bursts (629) --- High energy astrophysics (739))}

\section{Introduction} 
\label{sec:intro}


On 9 October 2022, at 13:16:59.99 universal time (UT) (hereafter $T_0$), the Gamma-ray Burst Monitor (GBM) on the Fermi spacecraft triggered and located the burst GRB 221009A \citep{GBM2022GCN.32636....1V}. It is also detected by other gamma-ray detectors, such as the Fermi Large Area Telescope (Fermi/LAT; \citealt{LAT2022GCN.32658....1P}), Konus-Wind \citep{KW2022GCN.32668....1F,KW2023}, INTEGRAL \citep{INTEGRAL2023arXiv230316943R}, Swift-BAT \citep{swift2022GCN.32688....1K}, GECAM-C \citep{Gecam2023,Yang2023ApJ...947L..11Y} and LHAASO \citep{LHAASO_2023}.
The exceptionally large fluence of this event saturated almost all gamma-ray detectors during the main burst. The event fluence of GRB
221009A was measured to be  $\rm 0.2~erg~cm^{-2}$ in the energy range of 10 to
1000 keV \citep{KW2023,Gecam2023,2023ApJ...952L..42L}, much higher than any previously detected GRB.

LHAASO observed  GRB 221009A at the epochs covering both the prompt emission
phase and the early afterglow in the TeV band, revealing the onset of afterglow emission in the TeV band as early as $T_0+230 ~{\rm s}$ and a peak at $\sim T_0+244~{\rm s}$ \citep{LHAASO_2023}. However, there is no evidence of afterglow emission at such early time at other wavelengths. In the optical to X-rays, the observations start too late (thousands of seconds after the GBM trigger), so they missed the early afterglow. In the hard X-ray to Mev $\gamma$-ray energies, as measured by GBM, the prompt emission is so bright that the afterglow emission is likely to be hidden by the prompt emission. The GBM observations show that the last discernible pulse peaking around $T_0+575$ s is followed by a long, smooth decay period \citep{2023ApJ...952L..42L}.  \citet{2023ApJ...952L..42L} suggest that this smooth decay is consistent with an afterglow origin. They obtain constraints for the peak of the afterglow at $\sim T_0+600 \rm~s$ by fitting the 10 keV extrapolated light curve with two \cite{Norris2005ApJ...627..324N} pulse models for the prompt emission
and a broken power-law model (BPL) for the afterglow emission. They tentatively identify  this  time as the
peak of onset of the solely afterglow emission.  
Additionally, \citet{2023ApJ...952L..42L}  obtain different constraints for the peak of the afterglow
by extrapolating the smooth emission of the time interval $T_0+597 - T_0+1467$~s back in time.  Requiring that
the extrapolated afterglow flux does not exceed the prompt emission flux, they place a limit of $t_{\rm peak}\ga 280~{\rm s}$ assuming the ISM medium for the afterglow emission. As noted in \citet{2023ApJ...952L..42L} ,  all
afterglow peak times mentioned above depend on the prompt emission pulse model and the BPL functional fits to the 10 keV
extrapolated light curve and therefore are model-dependent. 

 The apparent discrepancy between the onset time measured from the TeV afterglow and that inferred from the GBM analysis could be due to that the afterglow onset time in the GBM energy range is  earlier, but it is hidden by the bright prompt emission (as also argued in  \citet{2023ApJ...952L..42L} ). Indeed, 
in GRB 190114C, the afterglow emission is found to overlap with the highly variable prompt emission \citep{Ravasio2019A&A...626A..12R}.   
Motivated by this, we attempt to search for the afterglow emission at the dips of the GBM emission, where the prompt emission is at the lowest level.  We analyze the spectral evolution of the emission of GRB 221009A during  two intervals containing the dips  (i.e., intervals $T_{0}+[300-328] \rm ~s$ and $T_{0}+[348-378] \rm ~s$). We also analyze the spectral evolution of the emission after the last pulse peaking around $T_0+575$ s.  The data analysis and spectral results will be presented in \S 2. We find evidence of the transition from the prompt emission to afterglow at the two dips as well as during the period after the last pulse. Combining the afterglow data at the two dips with the late-time afterglow (after $T_0+660~{\rm s}$), we construct a light curve of the afterglow, which is presented in \S 3. In this section, we also discuss the modeling and implication of this afterglow component.  Finally, we give a summary in \S 4.

\section{Fermi-GBM data analysis} 
\label{sec:analysis}

Fermi-GBM is composed of twelve sodium iodide (NaI) detectors and two bismuth germanate (BGO) detectors \citep{GBM2009ApJ...702..791M}. The NaI detectors are sensitive to photons in the energy range from 8 keV to 900~keV, and the BGO detectors are sensitive in 200~keV-40~MeV. 
GRB 221009A was detected by Fermi-GBM beginning at the trigger time ($T_0$) and lasting until $T_0+1467$~s when it was occulted by the Earth.
In this work, we select two NaI detectors (NaI 4 and NaI
8) and one BGO detector (BGO 1) for preforming the data analysis, due to that they all have a smaller viewing angle, and the two NaI detectors stay within $60\degr$ until $T_0+1467$~s. 
We retrieved data files from the HEASARC online archive\footnote{\url{https://heasarc.gsfc.nasa.gov/FTP/fermi/data/gbm/daily/}} and the corresponding latest updated response matrix files (rsp2) from \citet{2023ApJ...952L..42L}.
Considering that  the spectra analysis of NaI data show deviations between the model and the data below 20 keV\footnote{In this work, we find that the NaI data also show deviations between the model and the data below 40 keV at $>600~\rm s$.}  \citep{2023ApJ...952L..42L} and  the presence of the Iodine K-edge at 33.17 keV \citep{2009ExA....24...47B}, we omit the NaI data below 40 keV in our analysis. Following  \citet{2023ApJ...952L..42L}, we also omit the BGO data below 400 keV in our analysis, due to  that the low energy incident photons that are affected by LAT are not adequately modeled in the detector response.   \citet{10MeV2023arXiv230316223E} reported that a highly significant narrow emission feature  around 10 MeV was found in $T_0+[280-320]$~s, therefore, we ignore the BGO data greater than 8000 keV in order to avoid its possible contamination in our analysis. 

\begin{figure*}
\includegraphics[angle=0,scale=0.63]{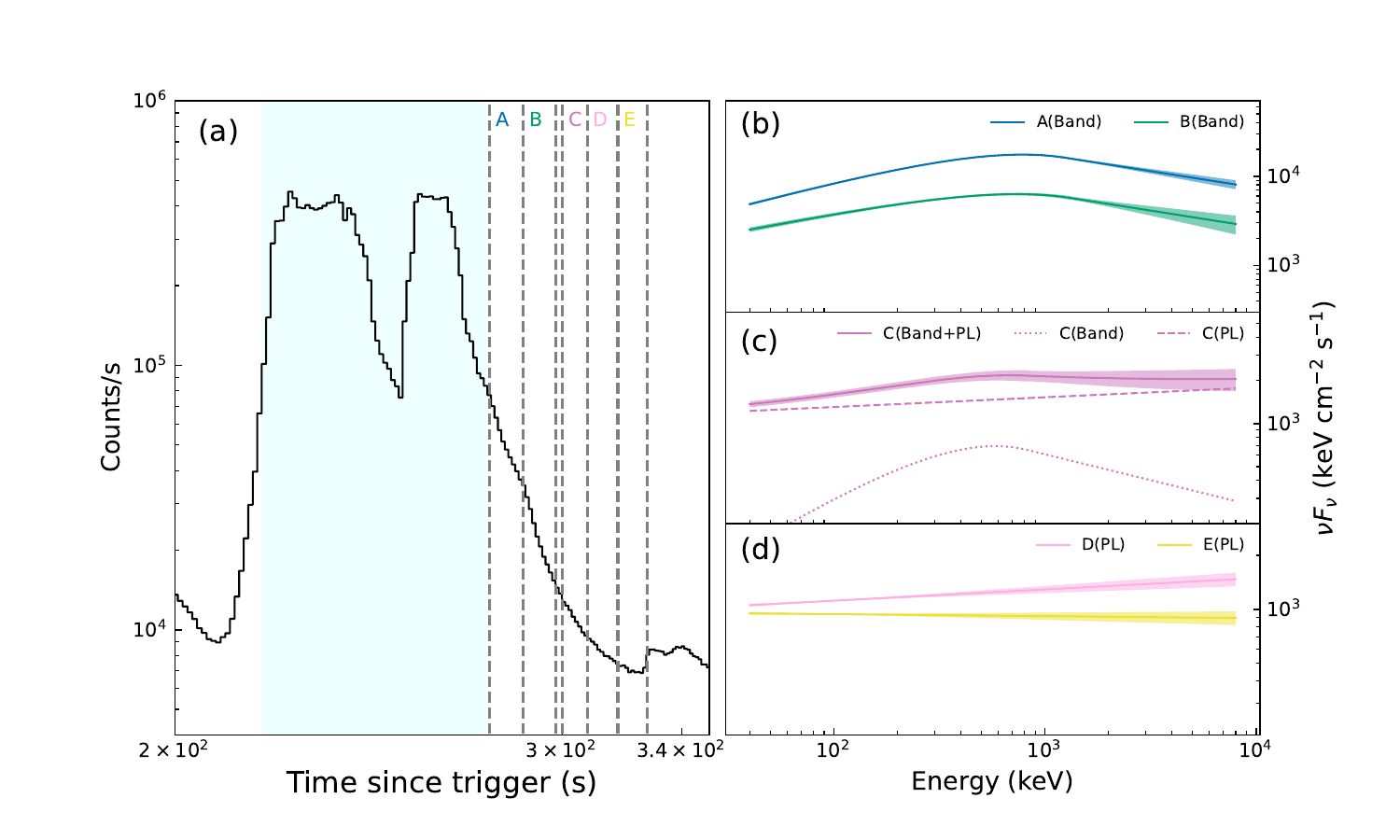}
\caption{(a): The light curve of GRB 221009A in 40--8,000 keV as seen by the NaI 4, NaI 8 and BGO 1 of Fermi-GBM during the interval $T_0+[288-328]\rm~s$. The light blue vertical shaded region shows the BTI ($T_0+[219-277]\rm~s$) of Fermi-GBM. The light curve is divided into five time slices (A-E, differentiated by the vertical dashed lines) for the spectral analysis. (b-d): The model spectra in $\nu F_\nu$ units for all five time slices.  The best-fit model for the  spectra in slices A and B is the Band function.
The best-fit model for the spectrum in slice c is Band+PL function.
The best-fit model for the  spectra in slices D and E is the PL function .
}
\label{lcsed1}
\end{figure*}

The GBM light curve, as shown in panels (a) of Figure \ref{lcsed1}, \ref{lcsed2} and \ref{lcsed3}, are derived from CSPEC type data\footnote{\url{https://fermi.gsfc.nasa.gov/ssc/data/access/gbm/}} of NaI 4, NaI 8 and BGO 1 in the energy range 40--8,000~keV. 
To model the background, we perform a standard polynomial fitting technique to the light curve over time intervals before and after emission episode. In this work, we select the same background time intervals as that shown in Figure 3 of \citet{2023ApJ...952L..42L}, during which  a fourth order polynomial fit  matches  the orbital background estimate  the best.

We extract the spectra for this burst with the public Fermi-GBM Data Tools Python software package (GDT; \citep{GDT2022}). 
We exclude the bad time intervals (BTIs, i.e., $T_{0}+[219-277]~\rm s$ \citep{Liu2023ApJ...943L...2L,2023ApJ...952L..42L} and $T_{0}+[508-514]~\rm s$ \citep{2023ApJ...952L..42L}).
We start the analysis at $T_{0}+278~{\rm s}$ to avoid the impact of pile-up, {and take 10~s width as a bin until $\rm T_{0}+508~s$.}
To assess the spectral evolution after the BTI  (i.e. after $T_{0}+514~\rm s$), we
extracted a sequence of 50~s or 100~s width  bins until $T_0+1467$~s when it was occulted by the Earth.

We pay special attention to three time intervals of the GBM observations during which the flux drops to a dip:
I)  $T_{0}+[278-328]~\rm s$,  II) $T_{0}+[328-378]~\rm s$  and III)  $T_0+[514-1467]~\rm s$  (see the left panels of Figures \ref{lcsed1}, \ref{lcsed2} and \ref{lcsed3}, where the count rate light curves around these time intervals are shown.). We divide each interval into several time slices for performing  spectral analysis, denoted with vertical dashed lines in the left panels of Figures \ref{lcsed1}, \ref{lcsed2} and \ref{lcsed3}. Throughout our spectral analyses, we test three spectral models, including a Band function (Band; \citet{band1993ApJ...413..281B}), a power-law (PL), and combinations of them (Band+PL).  The results of the spectral analysis of three selected time intervals are reported in  Table \ref{tab:specfit}.

To determine the best spectral model for the GBM data of GRB 221009A, we employ Bayesian information criterion (BIC; \citet{BIC1978AnSta...6..461S}) to compare different models. The BIC is defined as $\rm {BIC} = \chi^2 + k {\rm ln} N$, where $\chi^2$ is the PGstat statistic, $k$ and $N$ are the number of free parameters of the model and the number of data points, respectively. 
{ The details of PGstat statistic for the PL, Band and PL+Band models of all time intervals are shown in Table \ref{tab:specfit2} in the Appendix.}
For interval I ($T_{0}+[278-328]~\rm s$), as shown in Table \ref{tab:specfit}, we find that the spectra are best described by the Band function for slices of A and B { with high $\Delta_{\rm BIC}$ values compared with PL function or Band+PL models}\footnote{As suggested by \citet{BIC2017JCAP...01..005N}, the strength of the evidence against the model with the higher BIC value can be summarized as follows. (1) If $0<\Delta_{\rm BIC}<2$, there is no evidence against the higher BIC model; (2) if $2<\Delta_{\rm BIC}<6$, positive evidence against the higher BIC model is given; (3) if $6<\Delta_{\rm BIC}<10$, strong evidence against the higher BIC model is given; (4) if $\Delta_{\rm BIC}>10$, a very strong evidence against the higher BIC model is given.}. For  slice C ($T_{0}+[300-308]~\rm s$), we find that the combinations of a Band function and a PL provides the best fit with a difference of BIC between the Band function and Band+PL function being $\Delta_{\rm BIC}=8.70$.  Later on (i.e., during slices D and E), the spectra are best fitted by PL. This indicates that as the Band component decreases the PL component emerges. Since the prompt emission is usually characterized by a Band function while the afterglow is characterized by a PL, we may see a transition from the prompt emission to the underlying afterglow during this time interval. { Our result is consistent with the spectral fitting result of \citet{2023ApJ...952L..42L}, in which the authors find  that the Band+PL function (with PL index = $-1.916\pm0.009$ ) best fits the data during $T_0+[277.894-323.975]$~s.}

For interval II ($T_{0}+[328-378]~\rm s$),  we find that that the spectra are best described by the combinations of a Band function and a PL for slices of A and B. For slices C to E, the spectra are best fitted by a PL { (see  Table \ref{tab:specfit2} in the Appendix}). This indicates that the prompt emission component drops and the underlying afterglow component emerges at the dip. 

For interval III ($T_0+[514-1467]~\rm s$), we find that  the spectra are best described by the Band function for slices of A and B. For slices C, we find that the combinations of a Band function and a PL provides the best fit. After this,  the spectra in all slices are best fitted by a PL { (see Table \ref{tab:specfit2} in the Appendix}).  Coincidently, the light curve after slice B shows a remarkably smooth decay in time.  Both the spectral and temporal characteristics point to a transition from the prompt emission to the afterglow. The smooth decay after the last pulse has been noticed in the work of \citet{2023ApJ...952L..42L},  which also attributed the long, smooth decay period after the final pulse to the afterglow emission.
The light curve observed by $INTEGRAL/{\rm IBIS}$ in 200--2600~keV also shows that the afterglow emission begins to dominate at $\sim T_0+630~\rm s$ \citep{INTEGRAL2023arXiv230316943R}. 

\section{Afterglow light curve}
\label{sec:lcfitting}


We extract the flux of the PL component in the energy range of 50--100 keV and 200--400 keV from the spectral analysis in all slices and plot them in Figure \ref{lc1} (see the black data points). Adopting the reference time of the afterglow light curve at $T^*=T_0+226~{\rm s}$ \citep{LHAASO_2023}, we find that these flux points form a remarkable power-law decay in time (see the left panel of Figure \ref{lc1}). The PL fit ($F\propto t^{\alpha_{pl}}$) results in slopes of $\alpha_{pl}=-0.94\pm0.02$ and $-0.95\pm0.05$ for 50--100 keV and 200-400 keV, respectively. This result strongly supports that the PL components in the spectral analysis originate from the afterglow emission. As the first data point of this PL afterglow is at $T_0+300 $~s, the peak of afterglow emission of GRB 221009A should be before $T_0+300 $ s. 



Early TeV observations of GRB 221009A reveals a light curve steepening at $T_{\rm b}=T^*+ 670^{+230}_{-110}~{\rm s}$, which is interpreted as a jet break \citep{LHAASO_2023}. The combined Insight-HXMT and GECAM-C observations of the hard X-ray emission suggests a break in the light curve between $T^*+650\,{\rm s}$ and $T^*+1100\,{\rm s}$ (most likely at $T^*+\sim950$ s), consistent with the LHAASO result \citep{Gecam2023}. Motivated by this, we study whether a break in the afterglow light curve can be revealed from the GBM data. 

First, we compare a broken power-law model having two free slopes with a single power-law model in fitting the light curve of the afterglow emission in 50--100 keV. The best-fitting result gives a slope of $\alpha_1=-0.93\pm 0.02$ before the break and a slope of $\alpha_2=-1.82^{+0.42}_{-0.31}$ after the break and a break time of $t_b=T^*+1032.41^{+98.55}_{-129.66} ~{\rm s}$. Compared with the single power-law model, the broken power-law model has $\Delta \rm BIC=4.64$ ($\Delta \rm BIC$ is defined as $\rm BIC_2-BIC_1$, where $\rm BIC_1$ and $\rm BIC_2$ are the  $\rm BIC$ values for the power-law model and broken power-law model, respectively), indicating that there is no strong evidence against the broken power-law model. 


Then, we consider a  scenario in which the slopes before and after the break are constrained by the jet break physics \citep{Rhoads1999ApJ...525..737R,Sari1999ApJ...519L..17S}. We consider two possible cases: { A}) one is that the jet break is due to the edge effect and the light curve steepens by a power of $\Delta \alpha =3/4$ (i.e., $\alpha_2=\alpha_1-3/4$) for a constant-density medium; { B}) the other is that the sideways expansion effect of the jet is significant so that the post-break decay is $t^{-p}$ with $p\simeq 2$ (here we set $\alpha_2=-2$).  In case { A}, the best-fitting result gives a slope of $\alpha_1=-0.92\pm 0.03$ before the break and  and a break time at $t_b=T^*+1004.75^{+109.29}_{-116.11}~{\rm s}$. Compared with the single power-law model, the broken power-law model has $\rm \Delta BIC=1.63$.
In case { B}, the best-fitting result shows that $\alpha_1=-0.93\pm 0.02$, $t_b=T*+1055.15^{+81.80}_{-83.42} {\rm s}$ and $\rm \Delta BIC=1.71$. We also apply the same analysis to the light curves of the afterglow emission in 200--400~keV. The slopes and break times are all consistent with the case for the 50--100~keV emission within the uncertainty. The fitting results are plotted in the right panel of Figure \ref{lc1} and summarized in  Table \ref{tab:LC}.
The $\rm BIC$ values indicate that the jet break model and the single power-law model fit the data almost equally well. This indicates that a jet break can neither be ruled out nor be favored using the GBM data alone. Nevertheless, the slopes and break time in the jet breakh model are consistent with those found in the LHAASO data within the $1\sigma$ uncertainty. 
The fact that the two models can not be distinguished is understandable since the inferred break time is quite close to the last few points of the GBM data.

We can constrain the lower limit of the jet break time through a chi-squared test (following the $\chi^2_{N-n}$ distribution with $N-n=17$ degrees of freedom, where $N$ is the number of data points and $n$ is number of free parameters of the broken power-law model), assuming that the jet break is present. To obtain a useful constraint, we need to fix the slope before the jet break, so we take $\alpha_1=-0.93$ following the above analysis. For the jet break  scenario in which the light curve steepens by a power of $\Delta \alpha =3/4$, we find that the lower limits (at the $95\%$ confidence level) of the jet break time are $671.2~{\rm s}$ for the afterglow in 50--100~keV and $405.7~{\rm s}$ for the afterglow in 200--400~keV, respectively. For the jet break scenario in which the  post-break decay is $t^{-2}$, we find that the lower limits (at the $95\%$ confidence level) of the jet break time are $776.7~{\rm s}$ for the afterglow in 50--100~keV and  $522.2~{\rm s}$ for the afterglow in 200--400~keV, respectively. These lower limits are not in contradiction with the jet break time at TeV energies measured by LHAASO within the uncertainty. 


The hard X-ray afterglow emission measured by GBM should be produced by synchrotron emission of relativistic electrons accelerated in the forward shock. In this scenario, as the cooling frequency likely lies below the observed band, the expected photon index is $dN_{\gamma}/dE_\gamma\propto E_{\gamma}^{-(p+2)/2}$, where $p$ is the PL index of the electron energy distribution (i.e., $dN_e/dE_e\propto E_e^{-p}$). The observed photon index in 40--8,000 keV is about $-2.0$, implying that $p\simeq 2.0$, { which is also argued in  \citet{2023ApJ...952L..42L}}.
In the synchrotron afterglow model, the expected decay slope is $F\propto t^{-(2-3p)/4}\simeq t^{-1.0}$ for a spherical shock \citep{Sari1998ApJ...497L..17S}. Both the measured pre-break slope in the jet break model and the slope in the single PL decay model are  consistent with the expectation (see Table \ref{tab:LC}). 

\section{Summary}

Through the spectral analysis of the Fermi-GRB  data of GRB 221009A, we find clear evidence of afterglow emission during the dips of the  prompt emission pulses. The spectra show a transition from the Band function to a PL function during these periods. In the last discernible pulse (around $T_{0}+660$~s), we also find that the spectrum transits from the Band function to a PL function. During this period, the emerge of the afterglow emission is well consistent with light curve transition from a variable emission to the smooth decaying emission.
Remarkably, the afterglow flux at different intervals are smoothly connected in a single power-law, as shown in Figure \ref{lc1}. This suggests that the afterglow peak of GRB 221009A should be $t_{\rm peak}\la T_0+300 $~s. 

We also test if there is a jet break in the afterglow emission revealed from the GBM data, motivated by that a jet break is identified in the LHAASO TeV light curve. We find that  the jet break model can not be distinguished from the single power-law model using the  GBM data alone. This indicates that a jet break can neither be ruled out nor be favored using the GBM data alone. This is understandable since the inferred break time is quite close to the last GBM data.  Nevertheless, the slopes and break time in the jet break model are consistent with those found in the LHAASO data.  In future, combining the GBM data with other multi-wavelength data, one should be able to reach a conclusive result, which is, however, beyond the scope of the current work.

\begin{acknowledgments}
{ We thank the anonymous referee for valuable suggestions and} thank Stephen Lesage for providing us the latest updated response matrix files and thank Jun Yang for helpful discussion in GBM data analysis.
The work is supported by the  National Key R$\&$D Program of China under grant No. 2022YFF0711404, the NSFC under grants Nos. 12121003, 12203022 and U2031105, the China Manned Spaced Project (CMS-CSST-2021-B11) and the Natural Science Foundation of Jiangsu Province grant BK20220757.

\end{acknowledgments}



\bibliography{main}{}

\begin{thebibliography}{}
\expandafter\ifx\csname natexlab\endcsname\relax\def\natexlab#1{#1}\fi
\providecommand{\url}[1]{\href{#1}{#1}}
\providecommand{\dodoi}[1]{doi:~\href{http://doi.org/#1}{\nolinkurl{#1}}}
\providecommand{\doeprint}[1]{\href{http://ascl.net/#1}{\nolinkurl{http://ascl.net/#1}}}
\providecommand{\doarXiv}[1]{\href{https://arxiv.org/abs/#1}{\nolinkurl{https://arxiv.org/abs/#1}}}

\bibitem[{{An} {et~al.}(2023){An}, {Antier}, {Bi}, {Bu}, {Cai}, {Cao}, {Camisasca}, {Chang}, {Chen}, {Chen}, {Chen}, {Chen}, {Chen}, {Chen}, {Chen}, {Coughlin}, {Cui}, {Dai}, {Hussenot-Desenonges}, {Du}, {Du}, {Du}, {Fan}, {Frontera}, {Gao}, {Gao}, {Ge}, {Gong}, {Gu}, {Guan}, {Guo}, {Guo}, {Guidorzi}, {Han}, {He}, {He}, {Hou}, {Huang}, {Huo}, {Ji}, {Jia}, {Jiang}, {Kann}, {Klotz}, {Kong}, {Lan}, {Li}, {Li}, {Li}, {Li}, {Li}, {Li}, {Li}, {Li}, {Li}, {Li}, {Li}, {Li}, {Li}, {Liang}, {Liang}, {Liao}, {Lin}, {Liu}, {Liu}, {Liu}, {Liu}, {Liu}, {Liu}, {Liu}, {Lu}, {Lu}, {Lu}, {Luo}, {Luo}, {Ma}, {Ma}, {Ma}, {Ma}, {Maccary}, {Mao}, {Meng}, {Nie}, {Orlandini}, {Ou}, {Peng}, {Peng}, {Qiao}, {Qu}, {Ren}, {Shi}, {Shi}, {Song}, {Song}, {Su}, {Sun}, {Sun}, {Sun}, {Tan}, {Tan}, {Tao}, {Tuo}, {Turpin}, {Wang}, {Wang}, {Wang}, {Wang}, {Wang}, {Wang}, {Wang}, {Wang}, {Wang}, {Wang}, {Wang}, {Wang}, {Wang}, {Wang}, {Wen}, {Wu}, {Wu}, {Wu}, {Xiao}, {Xiao}, {Xiao}, {Xie}, {Xiong}, {Xiong}, {Xu}, {Xu}, {Xu}, {Xu}, {Xu}, {Xu},
  {Xue}, {Yang}, {Yang}, {Yang}, {Ye}, {Yi}, {Yi}, {Yin}, {You}, {Yu}, {Yu}, {Yu}, {Zeng}, {Zhang}, {Zhang}, {Zhang}, {Zhang}, {Zhang}, {Zhang}, {Zhang}, {Zhang}, {Zhang}, {Zhang}, {Zhang}, {Zhang}, {Zhang}, {Zhang}, {Zhang}, {Zhang}, {Zhang}, {Zhang}, {Zhang}, {Zhao}, {Zhao}, {Zhao}, {Zhao}, {Zhao}, {Zhao}, {Zhao}, {Zhao}, {Zheng}, {Zheng}, {Zhou}, {Zhou}, \& {Zhu}}]{Gecam2023}
{An}, Z.-H., {Antier}, S., {Bi}, X.-Z., {et~al.} 2023, arXiv e-prints, arXiv:2303.01203, \dodoi{10.48550/arXiv.2303.01203}

\bibitem[{{Band} {et~al.}(1993){Band}, {Matteson}, {Ford}, {Schaefer}, {Palmer}, {Teegarden}, {Cline}, {Briggs}, {Paciesas}, {Pendleton}, {Fishman}, {Kouveliotou}, {Meegan}, {Wilson}, \& {Lestrade}}]{band1993ApJ...413..281B}
{Band}, D., {Matteson}, J., {Ford}, L., {et~al.} 1993, \apj, 413, 281, \dodoi{10.1086/172995}

\bibitem[{{Bissaldi} {et~al.}(2009){Bissaldi}, {von Kienlin}, {Lichti}, {Steinle}, {Bhat}, {Briggs}, {Fishman}, {Hoover}, {Kippen}, {Krumrey}, {Gerlach}, {Connaughton}, {Diehl}, {Greiner}, {van der Horst}, {Kouveliotou}, {McBreen}, {Meegan}, {Paciesas}, {Preece}, \& {Wilson-Hodge}}]{2009ExA....24...47B}
{Bissaldi}, E., {von Kienlin}, A., {Lichti}, G., {et~al.} 2009, Experimental Astronomy, 24, 47, \dodoi{10.1007/s10686-008-9135-4}

\bibitem[{{Frederiks} {et~al.}(2022){Frederiks}, {Lysenko}, {Ridnaia}, {Svinkin}, {Tsvetkova}, {Ulanov}, {Cline}, \& {Konus-Wind Team}}]{KW2022GCN.32668....1F}
{Frederiks}, D., {Lysenko}, A., {Ridnaia}, A., {et~al.} 2022, GRB Coordinates Network, 32668, 1

\bibitem[{{Frederiks} {et~al.}(2023){Frederiks}, {Svinkin}, {Lysenko}, {Molkov}, {Tsvetkova}, {Ulanov}, {Ridnaia}, {Lutovinov}, {Lapshov}, {Tkachenko}, \& {Levin}}]{KW2023}
{Frederiks}, D., {Svinkin}, D., {Lysenko}, A.~L., {et~al.} 2023, \apjl, 949, L7, \dodoi{10.3847/2041-8213/acd1eb}

\bibitem[{{Goldstein} {et~al.}(2022){Goldstein}, {Cleveland}, \& {Kocevski}}]{GDT2022}
{Goldstein}, A., {Cleveland}, W.~H., \& {Kocevski}, D. 2022, Fermi GBM Data Tools: v1.1.1.

\bibitem[{{Krimm} {et~al.}(2022){Krimm}, {Barthelmy}, {Dichiara}, {Laha}, {Lien}, {Markwardt}, {Palmer}, {Parsotan}, {Sakamoto}, \& {Stamatikos}}]{swift2022GCN.32688....1K}
{Krimm}, H.~A., {Barthelmy}, S.~D., {Dichiara}, S., {et~al.} 2022, GRB Coordinates Network, 32688, 1

\bibitem[{{Lesage} {et~al.}(2023){Lesage}, {Veres}, {Briggs}, {Goldstein}, {Kocevski}, {Burns}, {Wilson-Hodge}, {Bhat}, {Huppenkothen}, {Fryer}, {Hamburg}, {Racusin}, {Bissaldi}, {Cleveland}, {Dalessi}, {Fletcher}, {Giles}, {Hristov}, {Hui}, {Mailyan}, {Malacaria}, {Poolakkil}, {Roberts}, {von Kienlin}, {Wood}, {Ajello}, {Arimoto}, {Baldini}, {Ballet}, {Baring}, {Bastieri}, {Gonzalez}, {Bellazzini}, {Bissaldi}, {Blandford}, {Bonino}, {Bruel}, {Buson}, {Cameron}, {Caputo}, {Caraveo}, {Cavazzuti}, {Chiaro}, {Cibrario}, {Ciprini}, {Orestano}, {Crnogorcevic}, {Cuoco}, {Cutini}, {D'Ammando}, {De Gaetano}, {Di Lalla}, {Di Venere}, {Dom{\'\i}nguez}, {Fegan}, {Ferrara}, {Fleischhack}, {Fukazawa}, {Funk}, {Fusco}, {Galanti}, {Gammaldi}, {Gargano}, {Gasbarra}, {Gasparrini}, {Germani}, {Giacchino}, {Giglietto}, {Gill}, {Giroletti}, {Granot}, {Green}, {Grenier}, {Guiriec}, {Gustafsson}, {Hays}, {Hewitt}, {Horan}, {Hou}, {Kuss}, {Latronico}, {Laviron}, {Lemoine-Goumard}, {Li}, {Liodakis}, {Longo}, {Loparco}, {Lorusso},
  {Lovellette}, {Lubrano}, {Maldera}, {Manfreda}, {Mart{\'\i}-Devesa}, {Mazziotta}, {McEnery}, {Mereu}, {Meyer}, {Michelson}, {Mizuno}, {Monzani}, {Morselli}, {Moskalenko}, {Negro}, {Nuss}, {Omodei}, {Orlando}, {Ormes}, {Paneque}, {Panzarini}, {Persic}, {Pesce-Rollins}, {Pillera}, {Piron}, {Poon}, {Porter}, {Principe}, {Rain{\`o}}, {Rando}, {Rani}, {Razzano}, {Razzaque}, {Reimer}, {Reimer}, {Ryde}, {S{\'a}nchez-Conde}, {Parkinson}, {Scotton}, {Serini}, {Sgr{\`o}}, {Sharma}, {Siskind}, {Spandre}, {Spinelli}, {Tajima}, {Torres}, {Valverde}, {Venters}, {Wadiasingh}, {Wood}, \& {Zaharijas}}]{2023ApJ...952L..42L}
{Lesage}, S., {Veres}, P., {Briggs}, M.~S., {et~al.} 2023, \apjl, 952, L42, \dodoi{10.3847/2041-8213/ace5b4}

\bibitem[{{LHAASO Collaboration}(2023)}]{LHAASO_2023}
{LHAASO Collaboration}. 2023, Science, 380, 1390, \dodoi{10.1126/science.adg9328}

\bibitem[{{Liu} {et~al.}(2023){Liu}, {Zhang}, \& {Wang}}]{Liu2023ApJ...943L...2L}
{Liu}, R.-Y., {Zhang}, H.-M., \& {Wang}, X.-Y. 2023, \apjl, 943, L2, \dodoi{10.3847/2041-8213/acaf5e}

\bibitem[{{Meegan} {et~al.}(2009){Meegan}, {Lichti}, {Bhat}, {Bissaldi}, {Briggs}, {Connaughton}, {Diehl}, {Fishman}, {Greiner}, {Hoover}, {van der Horst}, {von Kienlin}, {Kippen}, {Kouveliotou}, {McBreen}, {Paciesas}, {Preece}, {Steinle}, {Wallace}, {Wilson}, \& {Wilson-Hodge}}]{GBM2009ApJ...702..791M}
{Meegan}, C., {Lichti}, G., {Bhat}, P.~N., {et~al.} 2009, \apj, 702, 791, \dodoi{10.1088/0004-637X/702/1/791}

\bibitem[{{Norris} {et~al.}(2005){Norris}, {Bonnell}, {Kazanas}, {Scargle}, {Hakkila}, \& {Giblin}}]{Norris2005ApJ...627..324N}
{Norris}, J.~P., {Bonnell}, J.~T., {Kazanas}, D., {et~al.} 2005, \apj, 627, 324, \dodoi{10.1086/430294}

\bibitem[{{Nunes} {et~al.}(2017){Nunes}, {Pan}, {Saridakis}, \& {Abreu}}]{BIC2017JCAP...01..005N}
{Nunes}, R.~C., {Pan}, S., {Saridakis}, E.~N., \& {Abreu}, E. M.~C. 2017, \jcap, 2017, 005, \dodoi{10.1088/1475-7516/2017/01/005}

\bibitem[{{Pillera} {et~al.}(2022){Pillera}, {Bissaldi}, {Omodei}, {La Mura}, {Longo}, \& {Fermi-LAT team}}]{LAT2022GCN.32658....1P}
{Pillera}, R., {Bissaldi}, E., {Omodei}, N., {et~al.} 2022, GRB Coordinates Network, 32658, 1

\bibitem[{{Ravasio} {et~al.}(2019){Ravasio}, {Oganesyan}, {Salafia}, {Ghirlanda}, {Ghisellini}, {Branchesi}, {Campana}, {Covino}, \& {Salvaterra}}]{Ravasio2019A&A...626A..12R}
{Ravasio}, M.~E., {Oganesyan}, G., {Salafia}, O.~S., {et~al.} 2019, \aap, 626, A12, \dodoi{10.1051/0004-6361/201935214}

\bibitem[{{Ravasio} {et~al.}(2023){Ravasio}, {Sharan Salafia}, {Oganesyan}, {Mei}, {Ghirlanda}, {Ascenzi}, {Banerjee}, {Macera}, {Branchesi}, {Jonker}, {Levan}, {Malesani}, {Mulrey}, {Giuliani}, {Celotti}, \& {Ghisellini}}]{10MeV2023arXiv230316223E}
{Ravasio}, M.~E., {Sharan Salafia}, O., {Oganesyan}, G., {et~al.} 2023, arXiv e-prints, arXiv:2303.16223, \dodoi{10.48550/arXiv.2303.16223}

\bibitem[{{Rhoads}(1999)}]{Rhoads1999ApJ...525..737R}
{Rhoads}, J.~E. 1999, \apj, 525, 737, \dodoi{10.1086/307907}

\bibitem[{{Rodi} \& {Ubertini}(2023)}]{INTEGRAL2023arXiv230316943R}
{Rodi}, J., \& {Ubertini}, P. 2023, arXiv e-prints, arXiv:2303.16943, \dodoi{10.48550/arXiv.2303.16943}

\bibitem[{{Sari} {et~al.}(1999){Sari}, {Piran}, \& {Halpern}}]{Sari1999ApJ...519L..17S}
{Sari}, R., {Piran}, T., \& {Halpern}, J.~P. 1999, \apjl, 519, L17, \dodoi{10.1086/312109}

\bibitem[{{Sari} {et~al.}(1998){Sari}, {Piran}, \& {Narayan}}]{Sari1998ApJ...497L..17S}
{Sari}, R., {Piran}, T., \& {Narayan}, R. 1998, \apjl, 497, L17, \dodoi{10.1086/311269}

\bibitem[{{Schwarz}(1978)}]{BIC1978AnSta...6..461S}
{Schwarz}, G. 1978, Annals of Statistics, 6, 461

\bibitem[{{Veres} {et~al.}(2022){Veres}, {Burns}, {Bissaldi}, {Lesage}, {Roberts}, \& {Fermi GBM Team}}]{GBM2022GCN.32636....1V}
{Veres}, P., {Burns}, E., {Bissaldi}, E., {et~al.} 2022, GRB Coordinates Network, 32636, 1

\bibitem[{{Yang} {et~al.}(2023){Yang}, {Zhao}, {Yan}, {Wang}, {Zhang}, {An}, {Cai}, {Li}, {Li}, {Liu}, {Liu}, {Ma}, {Meng}, {Peng}, {Qiao}, {Shao}, {Song}, {Tan}, {Wang}, {Wang}, {Wen}, {Xiao}, {Xue}, {Yang}, {Yin}, {Zhang}, {Zhang}, {Zhang}, {Zhang}, {Zheng}, {Zheng}, {Xiong}, \& {Zhang}}]{Yang2023ApJ...947L..11Y}
{Yang}, J., {Zhao}, X.-H., {Yan}, Z., {et~al.} 2023, \apjl, 947, L11, \dodoi{10.3847/2041-8213/acc84b}

\end{thebibliography}
\bibliographystyle{aasjournal}



\begin{figure*}
\includegraphics[angle=0,scale=0.63]{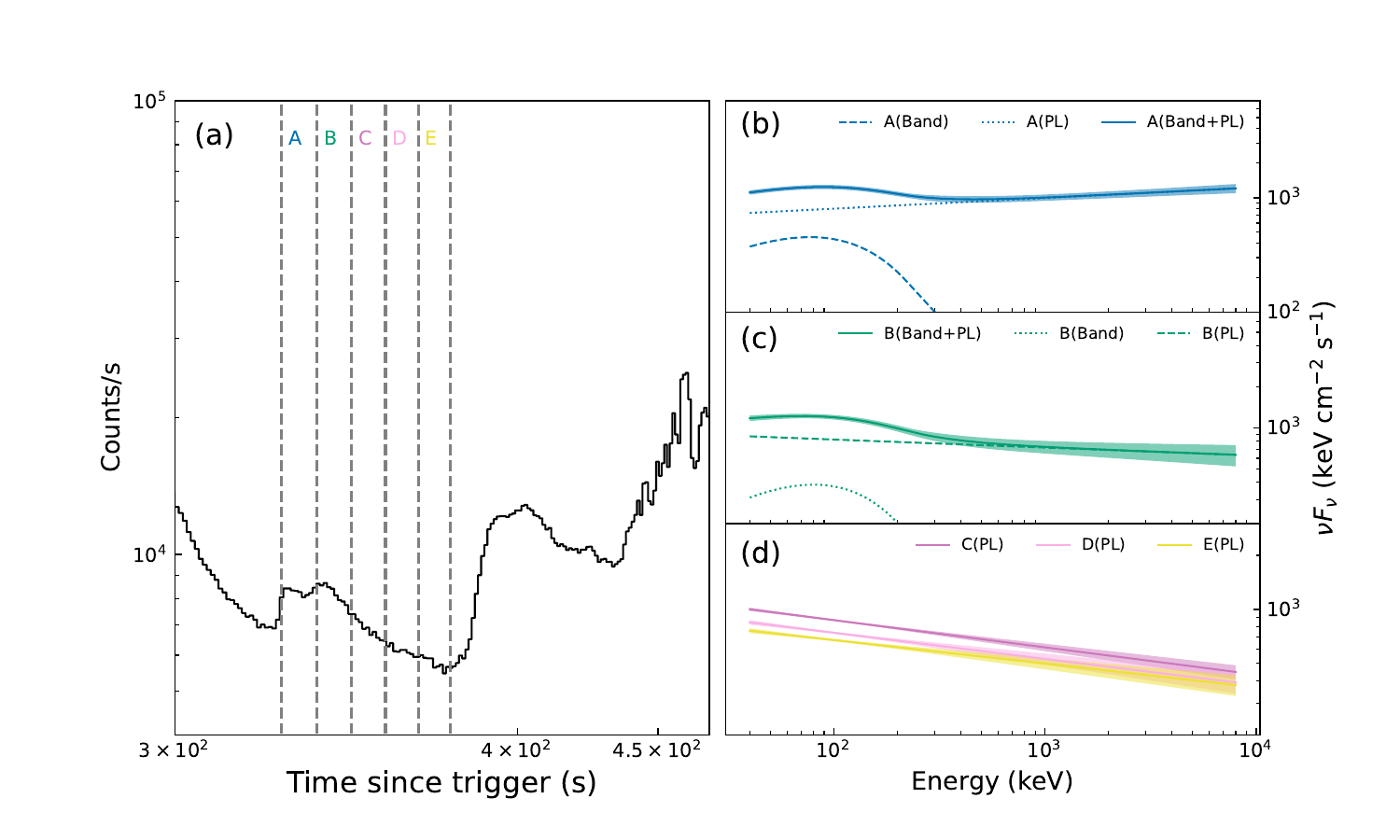}
\caption{Same as Figure \ref{lcsed1}, but for the time interval $T_0+[328-378]\rm~s$.  }
\label{lcsed2}
\end{figure*}

\begin{figure*}
\includegraphics[angle=0,scale=0.63]{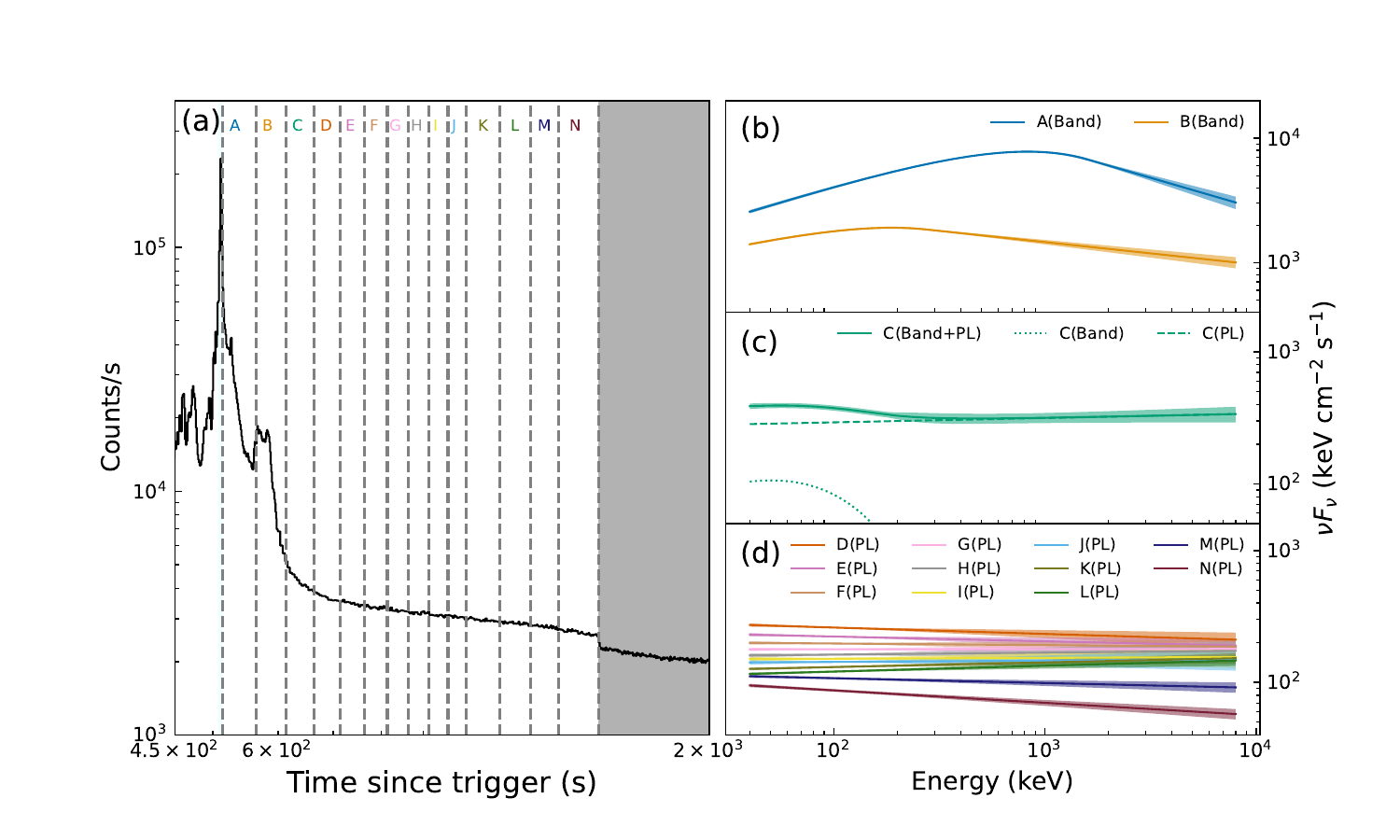}
\caption{Same as Figure \ref{lcsed1}, but for the time interval $T_0+[514-1467]\rm~s$.} 

\label{lcsed3}
\end{figure*}

\begin{figure*}
\includegraphics[angle=0,scale=0.68]{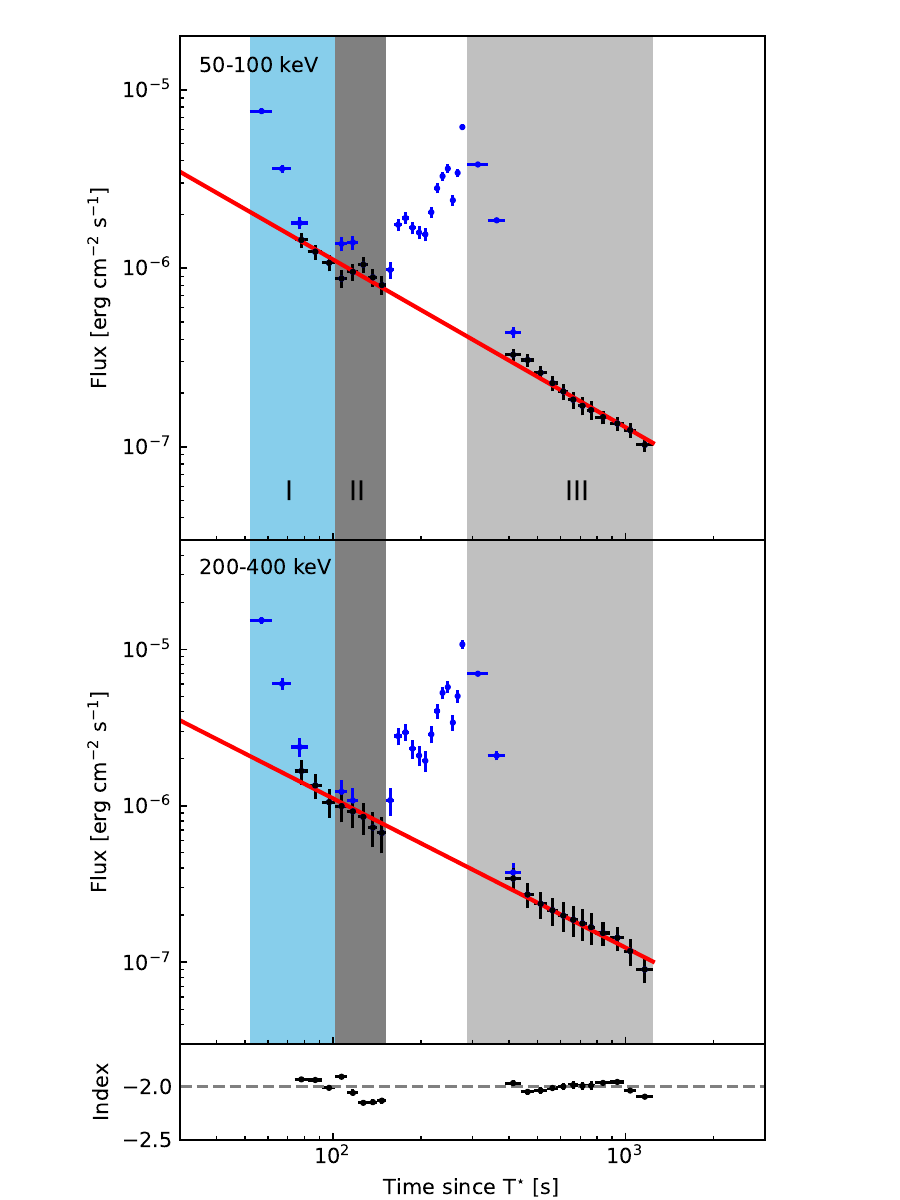}
\hspace{-20mm}
\includegraphics[angle=0,scale=0.68]{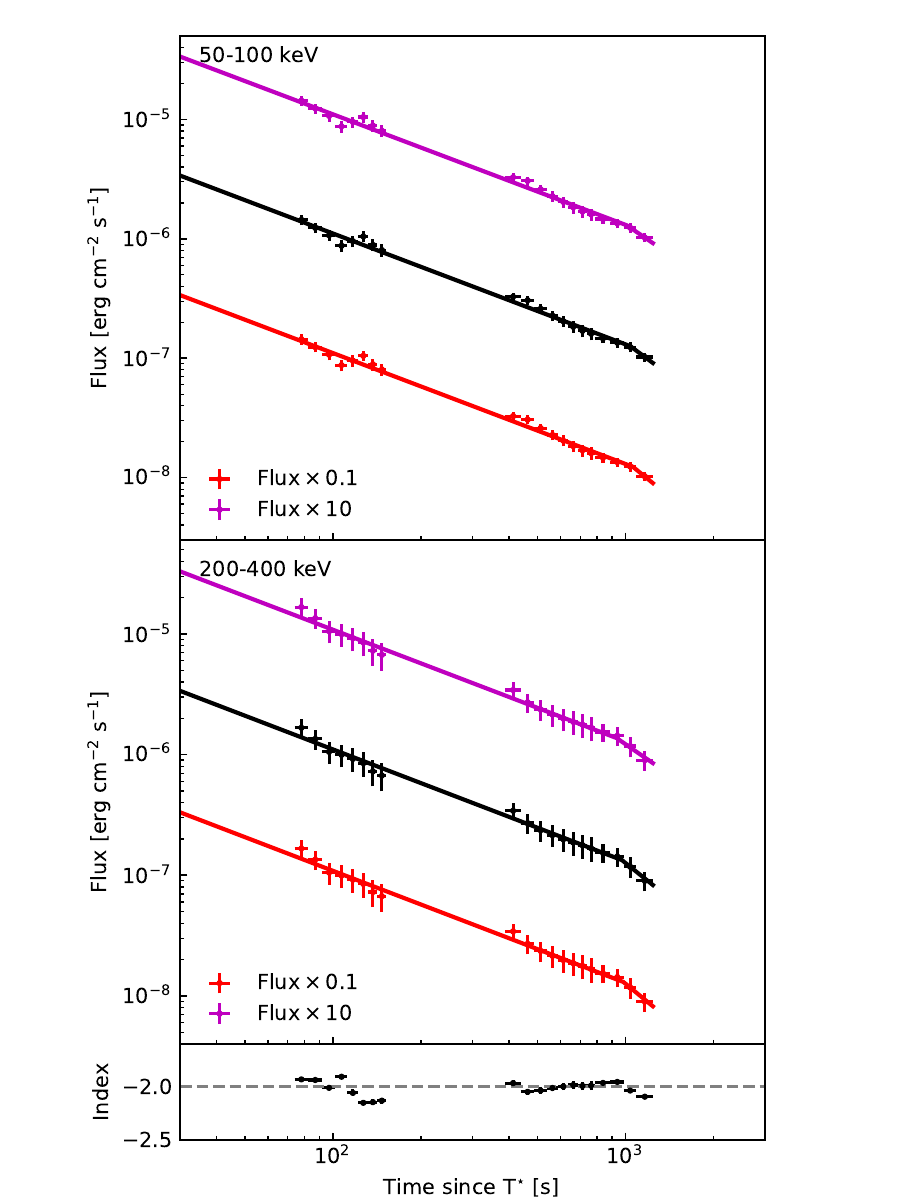}
\caption{Light curve data and modelling of GRB 221009A in 50--100 keV and 200--400 keV. 
{Left panel:} The three shaded regions represent the time intervals (I, II and III) that are selected for spectral analysis, as shown in Figure \ref{lcsed1}, Figure \ref{lcsed2} and Figure \ref{lcsed3}. The blue data points show the flux after fitting the spectra with the best model (Band, PL or Band+PL model), while the black data points show the flux of the PL component extracted from the spectral analysis. The red lines show the fitting of the black data points with a single power-law function, resulting in slopes of $-0.94\pm0.02$ and $-0.95\pm0.05$ for 50--100~keV and 200-400~keV, respectively.
{Right panel:} The afterglow component of GRB 221009A and modeling with the jet break models. The black lines show the fitting  with a broken power-law function with free $\alpha_1$ and $\alpha_2$, the magenta lines show the fitting with a broken power-law with free $\alpha_1$ and $\alpha_2=\alpha_1-3/4$, and the red lines show the fitting with a broken power-law with free $\alpha_1$ and $\alpha_2=-2.0$, respectively.
The bottom panels show the evolution of the PL photon index in 40--8,000 keV.
}
\label{lc1}
\end{figure*}

\begin{table}[ht!]
\caption{ Spectral fitting results  in  three time intervals.}
\scalebox{0.73}{
\begin{threeparttable} 
    \begin{tabular}{lccccccccc}
        \hline\hline
        Time interval & $\alpha$ & $\beta$ &  $E_p$ & Index & $\rm Flux_{50-100}$\tnote{*} & $\rm Flux_{200-400}$\tnote{*}  & PGstat/dof. & { Preferred model}  \\
        after $T_{0}$ (s) &  & & keV & &$\rm erg~cm^{-2}~s^{-1}$ & $\rm erg~cm^{-2}~s^{-1}$ \\ \hline
&&&&Interval I\\
\hline
A($278-288$) & $-1.37\pm0.01$ & $-2.37\pm0.06$ & $781.31\pm16.77$ & --   & $(7.58\pm0.29)\times10^{-6}$ & $(1.53\pm0.08)\times10^{-5}$ &632.80/267 & { Band} \\
B($288-298$)  & $-1.53\pm0.02$ & $-2.37\pm0.13$ & $733.73\pm34.06$ & --  & $(3.61\pm0.20)\times10^{-6}$&$(6.02\pm0.52)\times10^{-6}$ &444.42/267 & { Band}\\
($298-308$) &  $-1.77\pm0.21$ & $-2.11\pm0.25$ & $1172.97\pm316.22$ & --  & $(1.79\pm0.14)\times10^{-6}$ & $(2.37\pm0.33)\times10^{-6}$ &296.15/267 & { Band}\\
C($300-308$)  & $-1.08\pm0.10$ & $-2.36\pm0.30$  & $584.16\pm69.68$ & $-1.93\pm0.02$   & $(1.44\pm0.14)\times10^{-6}$ & $(1.67\pm0.31)\times10^{-6}$ &265.39/265 & { Band+PL}\\
D($308-318$)  & -- & -- & -- & $-1.94\pm0.02$  & $(1.24\pm0.12)\times10^{-6}$ & $(1.35\pm0.25)\times10^{-6}$ &330.24/269 & { PL}\\
E($318-328$)  & -- & -- & -- & $-2.01\pm0.02$  & $(1.07\pm0.11)\times10^{-6}$& $(1.05\pm0.22)\times10^{-6}$ &305.01/269 & {PL}\\
\hline
&&&&Interval II\\
\hline
A($328-338$) & $-0.91\pm0.03$  & $-4.01\pm0.84$ & $77.05^{+2.19}_{-2.15}$  & $-1.91\pm0.02$   & $(8.75\pm0.10)\times10^{-7}$ & $(9.95\pm2.12)\times10^{-7}$ &280.48/265 &{ Band+PL}\\
B($338-348$)  & $-0.89\pm0.03$  & $-4.00\pm0.66$ & $80.31\pm2.60$  & $-2.06^{+0.04}_{-0.02}$   & $(9.54\pm1.04)\times10^{-7}$ & $(9.21\pm2.04)\times10^{-7}$  &358.25/265 & { Band+PL}\\
C($348-358$) & -- & -- & --  & $-2.15\pm0.02$  & $(1.05\pm0.11)\times10^{-6}$ &  $(8.49\pm1.96)\times10^{-7}$  &338.13/269 & { PL}\\
D($358-368$) & -- & -- & --  & $-2.15\pm0.03$   & $(8.89\pm1.00)\times10^{-7}$ & $(7.27\pm1.81)\times10^{-7}$  &323.54/269 & { PL}\\
E($368-378$) & -- & --  &-- & $-2.13\pm0.03$  & $(8.04\pm0.96)\times10^{-7}$ & $(6.71\pm1.74)\times10^{-7}$  &354.94/269 & { PL}\\
\hline
&&&&Interval III\\
\hline
A($514-564$) & $-1.46\pm0.01$  & $-2.49^{+0.06}_{-0.07}$ & $  825.620^{+15.18}_{- 14.92}$  & --  & $(3.81\pm0.09)\times10^{-6}$ & $(6.98\pm0.25)\times10^{-6}$  &1275.51/267 & { Band}\\
B($564-614$) & $-1.60\pm0.01$  & $-2.18\pm0.03$ & $  187.99^{+3.86}_{- 3.75}$  & --  & $(1.86\pm0.06)\times10^{-6}$ & $(2.09\pm0.14)\times10^{-6}$&1058.44/267 & { Band}\\
C($614-664$) & $-1.18\pm0.02$  & $-4.11^{+2.10}_{-0.68}$ & $50.28^{+1.85}_{- 1.84}$  &$-1.97^{+0.03}_{-0.02}$  & $(4.37\pm0.31)\times10^{-7}$ & $(3.75\pm0.58)\times10^{-7}$&987.09/265 & { Band+PL}\\
D($664-714$) & -- & -- &   --  & $-2.05\pm0.03$  & $(3.06\pm0.26)\times10^{-7}$ & $(2.71\pm0.50)\times10^{-7}$ &1061.35/269 & { PL}\\
E($714-764$) & -- & -- &   --  & $-2.04\pm0.03$  & $(2.59\pm0.24)\times10^{-7}$ & $(2.36\pm0.46)\times10^{-7}$ &1178.75/269 & { PL}\\
F($764-814$) &  -- & -- &   --  & $-2.01\pm0.03$  & $(2.27\pm0.23)\times10^{-7}$ & $(2.15\pm0.44)\times10^{-7}$ &1032.81/269 & { PL}\\
G($814-864$) & -- & -- &   --  & $-2.00\pm0.03$  & $(2.04\pm0.22)\times10^{-7}$ & $(1.99\pm0.42)\times10^{-7}$ &1067.46/269 & { PL}\\
H($864-914$) & -- & -- &   --  & $-1.98\pm0.04$  &$(1.83\pm0.20)\times10^{-7}$  & $(1.87\pm0.41)\times10^{-7}$ &908.40/269 & { PL}\\
I($914-964$) & -- & -- &   --  & $-1.99\pm0.04$  & $(1.69\pm0.20)\times10^{-7}$ & $(1.77\pm0.40)\times10^{-7}$ &798.30/269 & { PL}\\
J($964-1014$) & -- & -- &   --  & $-1.99\pm0.04$  & $(1.60\pm0.19)\times10^{-7}$ & $(1.68\pm0.39)\times10^{-7}$ &825.59/269 & { PL}\\
K($1014-1114$) & -- & -- &   --  & $-1.96\pm0.02$  & $(1.46\pm0.13)\times10^{-7}$  & $(1.54\pm0.26)\times10^{-7}$ &1330.05/269 & { PL}\\
L($1114-1214$) & -- & -- &   --  & $-1.96\pm0.02$  & $(1.35\pm0.12)\times10^{-7}$ & $(1.43\pm0.25)\times10^{-7}$ &1042.46/269 & { PL}\\
M($1214-1314$) & -- & -- &   --  & $-2.04\pm0.02$  & $(1.24\pm0.12)\times10^{-7}$& $(1.18\pm0.23)\times10^{-7}$ &979.62/269 & { PL}\\
N($1314-1467$) & -- & -- &   --  & $-2.09\pm0.02$  & $(1.02\pm0.09)\times10^{-7}$& $(8.99\pm1.63)\times10^{-8}$ & 922.16/269 & { PL}\\
\hline\hline
\end{tabular}
    
\begin{tablenotes}   
        \footnotesize               
        { \item[*]  The $\rm Flux_{50-100}$ and $\rm Flux_{200-400}$ are obtained from Band function for preferred model is Band and PL function for the preferred model is PL (or Band+PL), respectively.}
\end{tablenotes} 
\end{threeparttable} 
}
\label{tab:specfit}
\end{table}

\begin{table}[ht!]
\caption{Fitting results of the afterglow light curve in 50-100 keV and 200-400 keV band with jet break models.}
\scalebox{0.75}{
\begin{threeparttable} 
    \begin{tabular}{lccc|cccccc}
        \hline\hline
      Energy band & \multicolumn{3}{c}{Power-law}  & \multicolumn{5}{c}{Broken Power-law} & \\ \hline
                  & $\alpha_{pl}$ &  $\chi^2/dof.$ &$\rm BIC_{1}$ & $\alpha_{1}$ & $\alpha_{2}$ & $t_{b}$(s) & $\chi^2/dof.$  & $\rm BIC_{2}$  & $\rm \Delta BIC$ \\
      \hline

     &  & & & $-0.93\pm0.02$ &  $-1.82^{+0.42}_{-0.31}$ & $1032.41^{+98.55}_{-129.66}$ & 10.86/16 & 22.84& 4.64  \\
  50-100 keV   & $-0.94\pm0.02$ & 12.21/18 & 18.20 & $-0.92\pm0.03$ & $\alpha_{2}=\alpha_{1}-3/4$ & $1004.75^{+109.29}_{-116.11}$ & 10.84/17 &19.83 & 1.63 \\
      &  & &  & $-0.93\pm0.02$ & $\alpha_{2}=-2.0$ & $1055.15^{+81.80}_{-83.42}$ & 10.92/17 &  19.91 & 1.71  \\
\hline
     &  & & & $-0.93\pm0.05$ &  $-1.95^{+1.00}_{-0.75}$ & $969.06^{+130.26}_{-188.80}$ &  2.18/16 &14.70 & 4.49  \\
  200-400 keV   & $-0.95\pm0.05$ & 4.22/18 &10.21 & $-0.92\pm0.05$ &  $\alpha_{2}=\alpha_{1}-3/4$ & $927.72^{+154.84}_{-149.15}$ & 2.83/17 &11.82 & 1.61  \\
      &  & & & $-0.93\pm0.05$ &  $\alpha_{2}=-2.0$ & $981.49^{+128.98}_{-121.48}$ & 2.73/17 &11.72 & 1.51  \\ 
\hline\hline
\end{tabular}
    
\end{threeparttable} 
}
\label{tab:LC}
\end{table}

\newpage
\appendix
{ 

In Figure \ref{spec2}, we show an example of count-rate spectrum  during a time interval ($\rm T_0 + 308 - T_0 + 318$~s) of interest in comparison with the three models. In addition, the values of PGstat statistic and $\rm \Delta BIC$ for the PL, Band and PL+Band models of all time intervals are shown in Table \ref{tab:specfit2}. Note that, in the time interval $\rm T_0+[378 - 388]$~s, the $\rm \Delta BIC_{PL}$ value is only 0.36, implying that there is no evidence against the Band model. Considering the light curve during this interval (as shown in Figure \ref{lcsed2})  shows a  rise, we think that this may be dominated by the prompt emission, and therefore the preferred model is selected to be the Band model.

\begin{figure*}
\includegraphics[angle=0,scale=0.41]{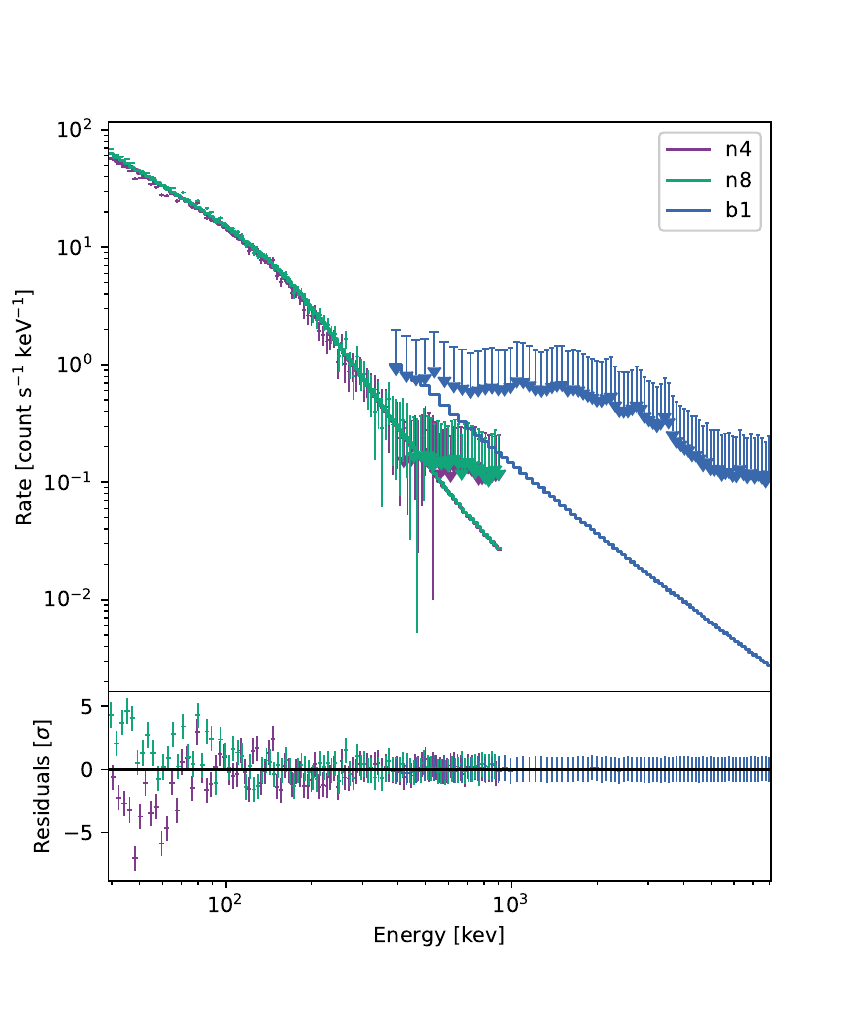}
\includegraphics[angle=0,scale=0.41]{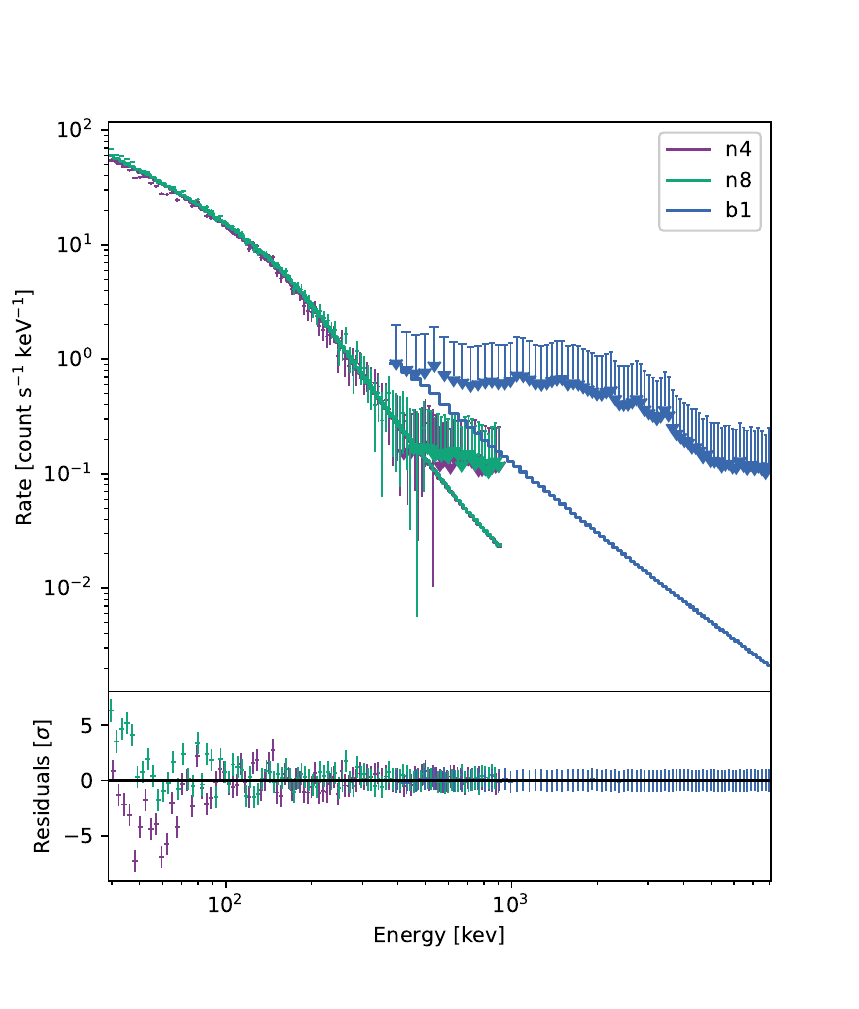}
\includegraphics[angle=0,scale=0.41]{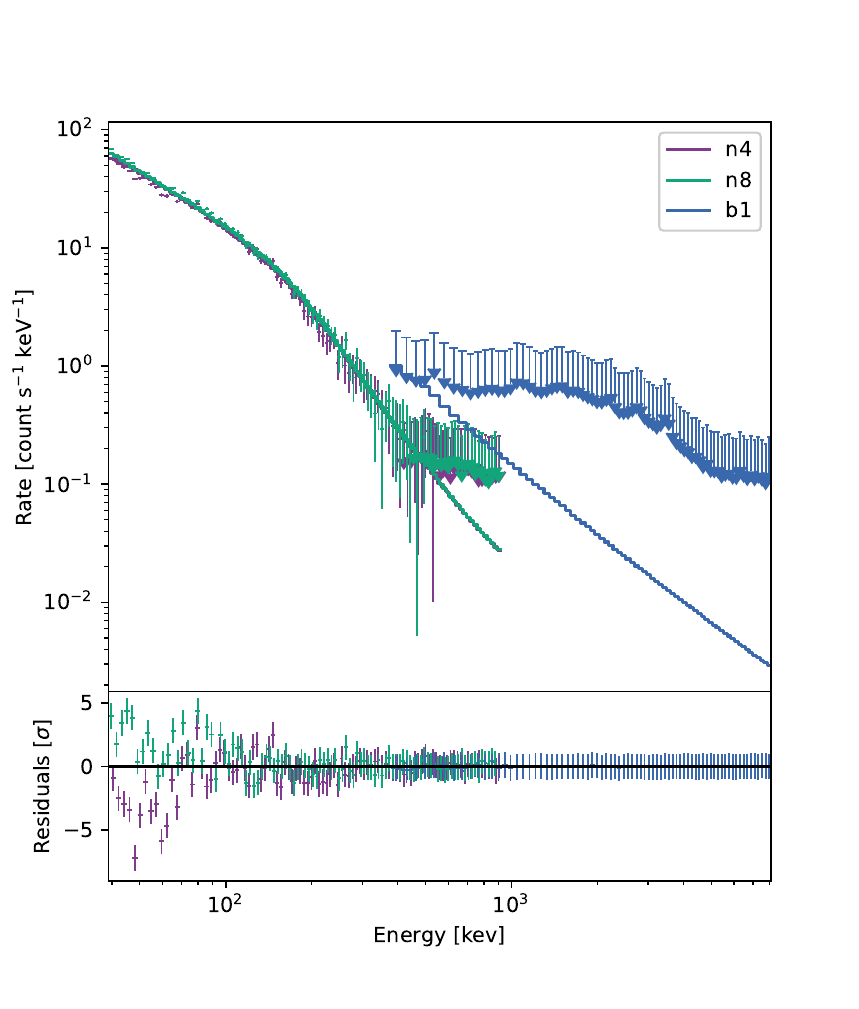}
\caption{{ An example of the count-rate spectrum  (during the interval $T_0+308$-$T_0+318$ s) in comparison with the PL (left panel), Band (middle panel) and Band+PL (right panel) models, respectively. The corresponding statistics for the fits are shown in Table 3 in Appendix.}
}
\label{spec2}
\end{figure*}

\begin{table}[ht!]
\caption{ { The values of PGstat statistics and $\Delta \rm BIC$ for the PL, Band and PL+Band model of all time intervals.}}
\scalebox{0.9}{
\begin{threeparttable} 
    \begin{tabular}{lcccccccc}
        \hline\hline
        Time interval & PGstat/dof. & PGstat/dof. & PGstat/dof. & $\Delta \rm BIC_{PL}$\tnote{$\dagger$} & $\Delta \rm BIC_{Band}$\tnote{$\dagger$} & $\Delta \rm BIC_{Band+PL}$\tnote{$\dagger$}  & Preferred model    \\
        after $T_{0}$ (s) & PL & Band & Band+PL &  &  &  \\ \hline
$278-288$ & 17173.77/269 & 632.80/267 & 870.21/265 & 16529.76  &  0  &  248.61  & Band   \\
$288-298$ & 1554.85/269 & 444.42/267 & 490.05/265 & 1099.22  &  0  &  56.83  &  Band \\
$298-308$  & 351.74/269 & 296.15/267 & 297.74/265 & 44.38  &  0  &  12.79 &  Band \\
$300-308$  & 296.43/269 & 285.29/267 & 265.39/265  & 8.63  &  8.70  &  0 &  Band+PL  \\
$308-318$  & 330.24/269  & 362.23/267 & 329.56/265  & 0  & 43.20  &  21.73  &  PL   \\
$318-328$  & 305.01/269 & 305.08/267 & 298.35/265 & 0  &  11.28  &  11.75 &  PL   \\
$328-338$  & 334.11/269 & 303.93/267 & 280.48/265 & 31.22  &  12.25  &  0  &  Band+PL  \\
$338-348$  & 440.86/269 & 398.23/267 & 358.25/265 & 60.20  &  28.78  &  0  &   Band+PL \\
$348-358$  & 338.13/269 & 333.50/267 & 338.13/265 & 0  &  6.58  &  22.41  &   PL \\
$358-368$ & 323.54/269 & 322.50/267 & 323.53/265 & 0  &  10.17  &  22.40 &   PL \\
$368-378$ & 354.94/269 & 356.44/267 & 346.65/265 & 0  &  12.71  & 14.12  &   PL \\
$378-388$ & 295.40/269 & 283.83/267 & 286.85/265 & 0.36  &  0  &  14.22  &   Band \\
$388-398$ & 834.46/269 & 308.39/267 & 302.02/265 &  514.86  &  0  &  4.83 &  Band  \\
$398-408$ & 936.85/269 & 335.18/267 & 355.43/265 &  590.46  &  0  &  31.45 &  Band  \\
$408-418$ & 643.77/269 & 330.05/267 & 329.81/265 & 302.51  &  0  &  10.96 &  Band  \\
$418-428$ & 552.77/269 & 320.37/267 & 319.58/265 & 221.19  &  0  &  10.41 &  Band  \\
$428-438$ & 518.97/269 & 322.26/267 & 320.69/265 & 185.50  &  0  &  9.63  &  Band  \\
$438-448$ & 802.55/269 & 380.78/267 & 380.44/265  & 410.56  &  0  &  10.86 &  Band  \\
$448-458$ & 1110.69/269 & 337.61/267 & 338.79/265 & 761.87  &  0  &  12.38 &  Band  \\
$458-468$ & 1722.38/269 & 441.86/267 & 440.23/265 & 1269.31  &  0  &  9.57 & Band  \\
$468-478$ & 1739.68/269 & 475.96/267 &475.51/265  & 1252.51  &  0  &  10.75  &  Band   \\
$478-488$ & 943.27/269 & 319.04/267 & 319.20/265 & 613.02  &  0  &  11.36 & Band   \\
$488-498$ & 1351.62/269 & 333.68/267 & 340.97/265 & 1006.73  &  0  &  18.49  &  Band  \\
$498-508$ & 3674.84/269 & 481.75/267 &534.21/265  & 3181.88  &  0  &  63.66 & Band   \\
$514-564$ & 8754.23/269 & 1275.51/267 & 1552.14/265 & 7467.51  &  0  &  287.83 & Band   \\
$564-614$ & 2107.49/269 &1058.44/267  & 1109.73/265 & 1037.84  &  0  &  62.49 & Band   \\
$614-664$ & 1012.37/269 & 1012.85/267 & 987.09/265  & 10.71  &  14.57  &  0 & Band+PL   \\
$664-714$ & 1061.35/269 & 1060.86/267 & 1057.92/265 & 0  &  10.72  &  18.98  &  PL  \\
$714-764$ & 1178.75/269 & 1698.63/267 & 1178.68/265 &  0  &  531.09  &  22.34 &  PL  \\
$764-814$ & 1032.81/269 & 1066.33/267 & 1032.50/265  & 0  &  44.73  &  22.10  &  PL  \\
$814-864$ & 1067.46/269 & 1080.44/267 & 1067.44/265  & 0  &  24.19  &  22.39  & PL   \\
$864-914$ & 908.40/269 & 916.15/267 & 901.63/265 & 0  &  18.96  &  15.64 &  PL  \\
$914-964$ & 798.30/269 & 803.67/267 & 798.18/265 & 0  &  16.58  &  22.29  & PL  \\
$964-1014$ & 825.59/269 & 828.18/267 & 822.71/265 & 0  &  13.8  &  19.53 &  PL  \\
$1014-1114$ & 1330.05/269 & 1330.08/267 & 1328.73/265 & 0  &  11.23  &  21.09  &  PL  \\
$1114-1214$ &1042.46/269  &1048.54/267 & 1028.10/265 & 0  &  17.29  &  8.05  &  PL  \\
$1214-1314$ & 979.62/269 & 979.79/267 &977.70/265  & 0  &  11.38  &  20.49 &  PL  \\
$1314-1467$ & 922.16/269 & 922.26/267 & 922.08/265 & 0  &  11.31  &  22.33  &  PL  \\
\hline\hline
\end{tabular}
\begin{tablenotes}   
        \footnotesize               
        \item[$\dagger$] The $\rm \Delta {BIC_{PL}}$, $\rm \Delta {BIC_{Band}}$ and $\rm \Delta {BIC_{Band+PL}}$ are defined as $\rm \Delta BIC_{PL}=BIC_{PL}-BIC_{Preferred \ model}$, $\rm \Delta BIC_{Band}=BIC_{Band}-BIC_{Preferred \ model}$ and $\rm \Delta BIC_{Band+PL}=BIC_{Band+PL}-BIC_{Preferred \ model}$. 
\end{tablenotes} 
\end{threeparttable} 
}
\label{tab:specfit2}
\end{table}

}


\end{document}